\documentclass[aps,prb,twocolumn,superscriptaddress,amsmath,amssymb]{revtex4}

\usepackage{overpic} 
\usepackage{braket}
\usepackage{graphicx}
\usepackage{xcolor}
\usepackage[pdf]{pstricks}
\usepackage[caption=false]{subfig}
\usepackage{bbold}
\usepackage{mathbbol}
\usepackage{bbm}
\usepackage{grffile}
\usepackage{epstopdf}
\graphicspath{{../plots/}}
\begin{document}


\title{Simulation of Charge Transport in Organic Semiconductors: A Time-Dependent Multiscale Method Based on Non-Equilibrium Green's Functions}




\author{S.\ Leitherer}
\altaffiliation{Present address: Department of Micro- and Nanotechnology, Technical University of Denmark,
\O rsteds Plads, 2800 Kgs. Lyngby, Denmark}
\affiliation{
Institute for Theoretical Physics and Interdisciplinary Center for Molecular Materials, \\
University Erlangen-N\"urnberg,
Staudtstr.\ 7/B2, D-91058 Erlangen, Germany
}
\author{C.\ M.\ J\"ager}
\affiliation{
Department of Chemical and Environmental Engineering,\\ University of Nottingham, University Park, NG7 2RD Nottingham, United Kingdom
}
\author{A.\ Krause}
\affiliation{
Computer-Chemie-Centrum and Interdisciplinary Center for Molecular Materials,\\ Department of Chemistry and Pharmacy, University Erlangen-N\"urnberg, N\"agelsbachstr.\ 25, 91052 Erlangen, Germany 
}
\author{M.\ Halik}
\affiliation{
Organic Materials \& Devices, Institute of Polymer Materials, Department of Materials Science, \\
University Erlangen-N\"urnberg,
Martensstr.\ 7, D-91058 Erlangen, Germany
}
\author{T.\ Clark}
\affiliation{
Computer-Chemie-Centrum and Interdisciplinary Center for Molecular Materials,\\ Department of Chemistry and Pharmacy, University Erlangen-N\"urnberg, N\"agelsbachstr.\ 25, 91052 Erlangen, Germany 
}
\author{M.\ Thoss}
\affiliation{
Institute for Theoretical Physics and Interdisciplinary Center for Molecular Materials, \\
University Erlangen-N\"urnberg,
Staudtstr.\ 7/B2, D-91058 Erlangen, Germany
}


\author{}
\affiliation{}

\date{\today}

\begin{abstract}
In weakly interacting organic semiconductors, static and dynamic disorder often have an important impact on transport properties. 
Describing charge transport in these systems requires an approach that correctly takes structural and electronic fluctuations into account. 
Here, we present a multiscale method based on a combination of molecular dynamics simulations, 
electronic structure calculations, and 
a transport theory that uses time-dependent non-equilibrium Green's functions. 
We apply the methodology to investigate the charge transport in  C$_{60}$-containing self-assembled monolayers (SAMs), which are used in organic field-effect transistors.
\end{abstract}


\maketitle


\textit{Introduction}.
Understanding the mechanisms of charge transport in organic semiconductors is both of fundamentals interest in condensed matter physics and a prerequisite for applications, which range from solar cells, organic light emitting devices or sensors to organic field-effect transistors (FETs).
For example, self-assembled monolayer field-effect transistors (SAMFETs),  \cite{sekitani2,schmaltz} containing thin films of $\pi$-conjugated molecules as semiconductor material
provide a promising platform for low-cost and flexible electronics.
In organic semiconductor materials, the structure is formed by molecules that are linked by weak van der Waals interactions.
In contrast to inorganic solids with highly periodic, rigid lattices, organic semiconductors often represent conformationally flexible systems, exhibiting a high degree of static and dynamic disorder.

Different theories have been set up to describe charge transport in organic semiconductors (for an overview see the Reviews \onlinecite{troisi11,Fratini16} and references therein).
While the short mean-free paths in the structures suggest hopping transport to be dominant, band-like transport has also been observed, indicated by a decrease of the mobility with increasing temperatures. 
In general, the existence of dynamic disorder requires a transport approach 
that takes different conformations and the mutual influence of structural and electronic properties into account.\cite{McMahon}
This can be achieved by combining molecular dynamics (MD) simulations, electronic structure calculations, and transport theory in a multiscale fashion, thus facilitating transport simulations without a priori assumptions about the dominant transport mechanism.\cite{Kubar13,popescu,Kubar16} 

\begin{figure}
\includegraphics[scale=0.72]{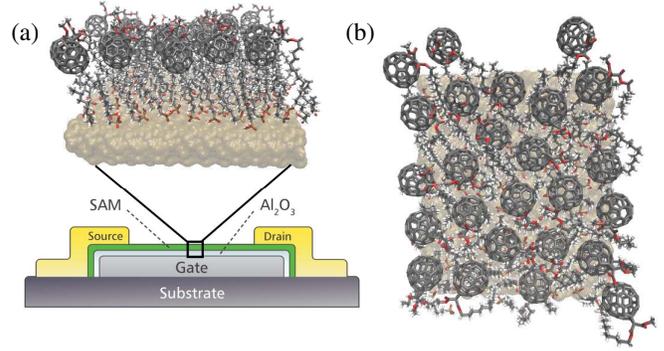}
 	 \caption{(a) Scheme of the SAMFET device containing the C$_{60}$-based SAM, 
 	 (b) snapshot of the SAM (top view), consisting of 75 C$_{10}$-PA and 25 C$_{60}$-C$_{18}$-PA molecules on AlO$_x$.}
 	 \label{fig:001}
\end{figure}

Within this methodological framework, we present here an efficient approach to study charge transport in organic semiconductors. 
We consider molecular structures, which, due to the influence of thermal fluctuations, exhibit rapidly oscillating electronic parameters,
in particular on-site energies and inter-site couplings.
To incorporate these fluctuations correctly, we employ a time-dependent transport approach based on nonequilibrium Green's function (NEGF) theory.\cite{croy,popescu2016} This method was
previously applied to study charge transport through DNA,\cite{Kubar13,Xie2013,popescu,chen2014} and has recently been extended to account for charge relaxation and electric field effects.\cite{Kubar16}
Here, we apply the methodology in a different setting to study charge transport in significantly larger organic structures,
in particular C$_{60}$-based SAMs used in FETs \cite{schmaltz2014,bauer2015} (cf.\ Fig.\ref{fig:001}). 
In Ref.\ \onlinecite{Leitherer14}, we have investigated charge transport in such SAMs based on a simpler methodology, 
which uses Landauer transport theory\cite{Landauer57} for selected structural snapshots along a molecular dynamics trajectory and time-averaging to obtain the electrical current. 
As has been shown previously,\cite{popescu} such an approach may fail for systems with fast-fluctuating electronic parameters.


\textit{Methods}.
The theoretical methodology we use to simulate charge transport in 
organic semiconductors consists of three steps: (i) characterization of the molecular structure using MD simulations, 
(ii) determination of the electronic structure, and (iii) charge transport calculations based on NEGF theory. 
%
%
%
%
%
Steps (i) and (ii) result in a time-dependent Hamiltonian describing the semiconductor,
given by $H_S(t)=\sum_{n}\tilde\epsilon_{n}(t)c_n^{\dagger}c_n+\sum_{n\ne m}\Delta_{nm}(t)c_n^{\dagger}c_m$,
where $\tilde\epsilon_{n}(t)$ are the time-dependent energies of the single-particle states $|\psi_{n}\rangle$, representing atomic orbitals in the system, 
and $\Delta_{nm}(t)$ are the time-dependent couplings between them, $c_n^{\dagger}$ and $c_n$ are creation and annihilation operators for the single-particle states employed. 
The semiconductor system is connected to left and right electrodes denoted by $\alpha=l,r$, which act as electron reservoirs (see below).


Based on this modeling, charge transport is described using time-dependent (TD) NEGF theory employing the propagation scheme presented in Ref.\ \onlinecite{croy}.
Thereby, the time-evolution of the reduced single-electron density matrix $\rho_{S;nm}(t)=\text{Tr}_S\left\{ \rho_S(t) c_m^{\dagger}c_n \right\}$ of the semiconductor is given by
\begin{align}
 i\frac{\partial}{\partial t}\rho_S(t) =[H_S(t),\rho_S(t)]+i\sum_{\alpha\in {l,r}}[\Pi_{\alpha}(t)+\Pi^{\dagger}_{\alpha}(t)],
 \label{eqn:eomrho}
\end{align}
with the current matrices
\begin{align}
\Pi_{\alpha}(t) = \int_{-\infty}^t dt_1\left[ G^>(t,t_1)\Sigma_{\alpha}^<(t_1,t)-G^<(t,t_1)\Sigma_{\alpha}^>(t_1,t)\right],
\label{eqn:auxcurr}
\end{align}
and the lesser/greater Green's functions $G^{\lessgtr}$ and selfenergies $\Sigma^{\lessgtr}$.
The former are defined as $G^{<}_{nm}(t,t')=i\langle c^{\dagger}_m(t')c_n(t)\rangle$,
with $G^{<}(t,t')=G^{>}(t',t)$. The reduced density matrix is related to the time-diagonal components of the lesser Green's function via $\rho_{S;nm}(t)=-iG^{<}_{nm}(t,t)$.

In the following, the wide-band approximation (WBA) is invoked, where the density of states in the electrodes is assumed to be energy-independent. 
Furthermore, explicit time-dependencies of the chemical potentials and of the electrode-molecule coupling are neglected.
With these assumptions, the lesser and greater self-energies can be written in an energy-resolved form\cite{Haug98}
\begin{align}
 \Sigma_{\alpha}^{\lessgtr}(t,t_1) =\pm i \int_{-\infty}^{\infty}\frac{dE}{2\pi}f_{\alpha}(\pm\beta(E-\mu_{\alpha}))e^{-iE(t-t_1)} \Gamma_{\alpha}.
 \label{eqn:enself}
\end{align}
Thereby, $\Gamma_{\alpha}$ denotes the spectral density in lead $\alpha$, which within the WBA is constant,
$f_{\alpha}$ the Fermi function for the electrons in the
left/right lead, $\beta =1/k_B T$, where $k_B$ is the Boltzmann constant and $T$ the electrode temperature.  
The integral in Eq.\ (\ref{eqn:enself}) can in general not be solved analytically. 
An auxiliary mode expansion (Pad\'{e} expansion \cite{Pade2010,Pade2011}) of the Fermi distribution is employed to transform the integral into a sum over $N_F$ poles
\begin{align}
f_{\alpha}(\beta(E-\mu_{\alpha}))
\approx \frac{1}{2}-\frac{1}{\beta}\sum_{p=1}^{N_F} \left(\frac{\kappa_p}{E-\chi^+_{\alpha p}}+\frac{\kappa_p}{E-\chi^-_{\alpha p}}\right).
 \label{eqn:fermipoles}
\end{align}
Thereby, $\chi^{\pm}_{\alpha p}=\mu_{\alpha}\pm i x_p/\beta$, where $x_p$ denotes the poles of the expansion, $\kappa_p$ are the Pad\'{e} coefficients,
and the chemical potentials $\mu_{\alpha}$ of the left/right electrode for a symmetric drop of the bias voltage $V$ around the Fermi energy $E_F$ are given by $\mu_{\alpha}=E_F\pm \frac{eV}{2}$.

With these assumptions, the current matrices $\Pi_{\alpha}(t)$ assume the form 
\begin{align}
 \Pi_{\alpha}(t)=\frac{1}{4}(\mathbbm{1} -2\rho_S(t))\Gamma_{\alpha}+\sum_{p=1}^{N_F} \Pi_{\alpha p}(t),
 \label{eqn:pi}
\end{align}
where $\Pi_{\alpha p}(t)$ are auxiliary current matrices, which obey the equation of motion 
\begin{align}
i\frac{\partial}{\partial t}\Pi_{\alpha p}(t) = \frac{\kappa_p}{\beta}\Gamma_{\alpha}+\left(H_S(t) -\frac{i}{2}\Gamma-\chi_{\alpha}^+\mathbbm{1}\right)\Pi_{\alpha p}(t),
\label{eqn:eompi}
\end{align}
with $\Gamma=\sum_{\alpha}\Gamma_{\alpha}$ and the initial condition $\Pi_{\alpha p}(t_0)=0$. 
The current from electrode $\alpha$ into the system is given by
\begin{align}
 I_{\alpha}(t) =\frac{2e}{\hbar}\operatorname{Re}\text{Tr}\left\{ \Pi_{\alpha}(t) \right\}, 
\end{align}
resulting in the net current $I(t)=\left[I_l(t)-I_r(t)\right]/2$.
 

\begin{figure*}[t]
\includegraphics[scale=0.75]{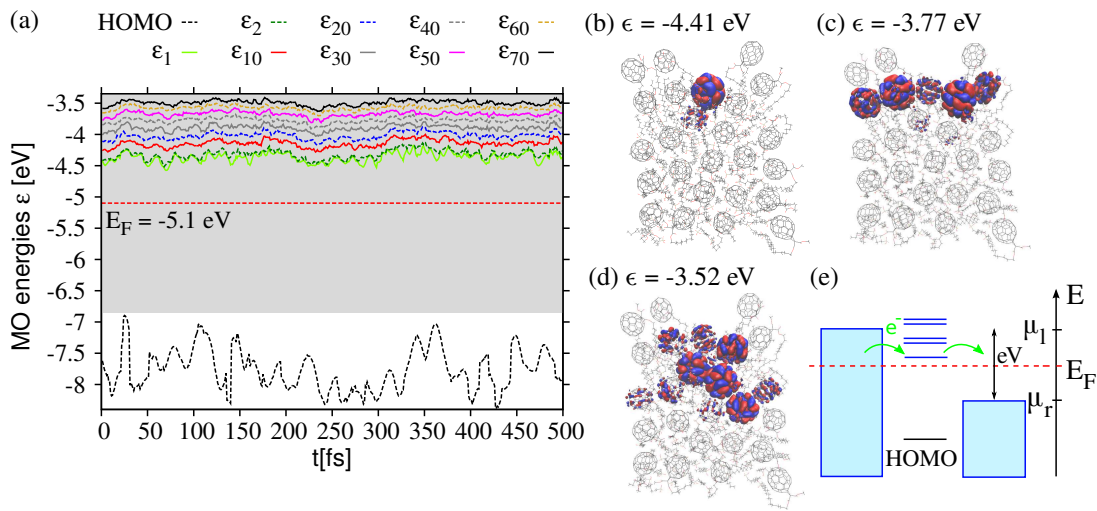}
 \caption{ (a) Time-resolved MO energies of the SAM. Shown are the energies of the HOMO and of unoccupied MOs.
 The red dashed line represents the Fermi level, $E_F=-5.1$ eV. The grey shaded area is the range of states which are located in the transport window when a bias voltage of 3.5 V is applied. (b-d) Examples of MOs from the unoccupied spectrum.
 (e) Energy level scheme of the device with the levels of the SAM coupled to the continuous spectrum of levels of the electron reservoirs in the electrodes,
 with temperature $T$ and chemical potentials $\mu_{\alpha}=E_F\pm \text{eV}/2$, where $V$ is the applied bias and $\alpha =l,r$. 
 }
   \label{fig:004}
\end{figure*}



\textit{Results and Discussion}.
We have employed the method presented above to study charge transport in C$_{60}$-based SAMs, which in the experiments\cite{tbauer} that inspired our theoretical studies 
are arranged in a SAMFET device as schematically shown in Fig.\ \ref{fig:001} (a).
The SAM is separated from the aluminum gate electrode at the bottom by a tiny AlO$_x$ layer. Lithographically 
patterned gold, placed on top of the SAM, serves as source and drain electrodes.\cite{jaeger}
The SAM is formed by fullerene-functionalized 
octadecyl-phosphonic acids (PAs) (in the following denoted by 
C$_{60}$-C$_{18}$-PA) and C$_{10}$-PA in a stoichiometric ratio of 1:3. 
The alkyl chains of the SAM, together with AlO$_x$, build the dielectric
of the device.\cite{tbauer}
The semiconducting C$_{60}$ head groups of the functionalized
 PA in the SAM form the active transistor channel in the device.
We focus on charge transport within 
the SAM; hence the influence of a gate potential and 
the AlO$_x$ layer is not taken into account. 
The basic unit representing the SAM is depicted in Fig.\ \ref{fig:001} (b). 
It comprises 25 C$_{60}$-C$_{18}$-PAs, mixed with 75 C$_{10}$-PAs.
The coupling to the gold electrodes is described implicitly using self-energies determined by the spectral density  $\Gamma_{\alpha}$. 
In accordance with the structure of the SAM, we use a model for the spectral density, where the matrix elements of $\Gamma_{\alpha}$, represented in a local basis of atomic orbitals,
are given by $(\Gamma_{\alpha})_{nn}=1$ eV
for orbitals $n$ corresponding to the outermost hexagon of carbon atoms 
of the C$_{60}$ head groups at the left and right boundary of the SAM
and $(\Gamma_{\alpha})_{nn'}= 0$ otherwise. This value is a reasonable choice 
for molecule-gold contacts. \cite{Bilan12,markussen11}

The conformational sampling of the SAM is based on classical atomistic 
MD simulations described in detail 
previously.\cite{jaeger,Leitherer14} 
Briefly, the AlO$_x$ surface was equilibrated prior to 
depositing PAs using an interatomic potential model 
parameterized by Sun et al. \cite{sunj} 
The parameters for the phosphonates are based on the general Amber force field (GAFF) \cite{wangj} and the MD simulations 
were performed with the program DL-POLY.\cite{todorovsmith} 
Following the MD simulations, the AlO$_x$ substrate was removed and for the molecular structure thus obtained, 
the electronic structure was determined for each snapshot of the MD trajectory by semiempirical molecular orbital (MO) calculations using the restricted Hartree-Fock formalism and the AM1 Hamiltonian.\cite{dewar} All semiempirical MO calculations were performed using 
the parallel EMPIRE program.\cite{empire}


We study the dynamics of the SAM after a simulation time of 100 ns, where the structure of the system is fully equilibrated. 
The analysis shows that after equilibration, there is no large-amplitude motion of the molecules in the SAM, however, there are significant thermal fluctuations.
These result in an explicitly time-dependent electronic structure of the SAM.
In order to take these rapid fluctuations into account correctly, the electronic structure is resolved with step size of 1 fs.
The spectrum of MO energies $\epsilon_j (t)$ of the SAM over a time span of 500 fs is displayed in Fig.\ \ref{fig:004} (a). 
Shown are the energies of the highest occupied molecular orbital (HOMO) and the lowest unoccupied molecular orbitals (LUMO), 
where $\epsilon_1$ corresponds to the LUMO of the SAM, $\epsilon_2$ to LUMO+1 etc.
Next to $\epsilon_1$ and $\epsilon_2$, the unoccupied MO energies $\epsilon_{10}-\epsilon_{70}$ are depicted in decimal steps, revealing a dense spectrum.

The frontier orbitals of the SAM are strongly localized due to the pronounced disorder in the system. 
A detailed analysis reveals that the occupied states are mainly localized on the anchor groups, while the lowest unoccupied states are localized on the  C$_{60}$ head groups.
Fig.\ \ref{fig:004} (b-d) shows several MOs from the unoccupied part of the spectrum, localized on few fullerenes in the SAM. 
The Fermi energy is set to the work function of gold ($E_F=-5.1$ eV) and is significantly closer to the unoccupied part of the spectrum.
The grey shaded area in Fig.\ \ref{fig:004} (a) indicates the energy range of electronic levels relevant for transport through the SAM for a voltage of 3.5 V, 
as defined by the symmetric voltage shift $\mu_{\alpha}=E_F\pm \text{eV}/2$. 
Despite strong fluctuations, the HOMO remains far away from the Fermi level. 
Therefore, only the unoccupied energy levels are relevant for transport and are taken into account in the calculations. This transport scenario is illustrated in Fig.\ \ref{fig:004} (e). 


\begin{figure}[t]
\centering
     \includegraphics[scale=0.72]{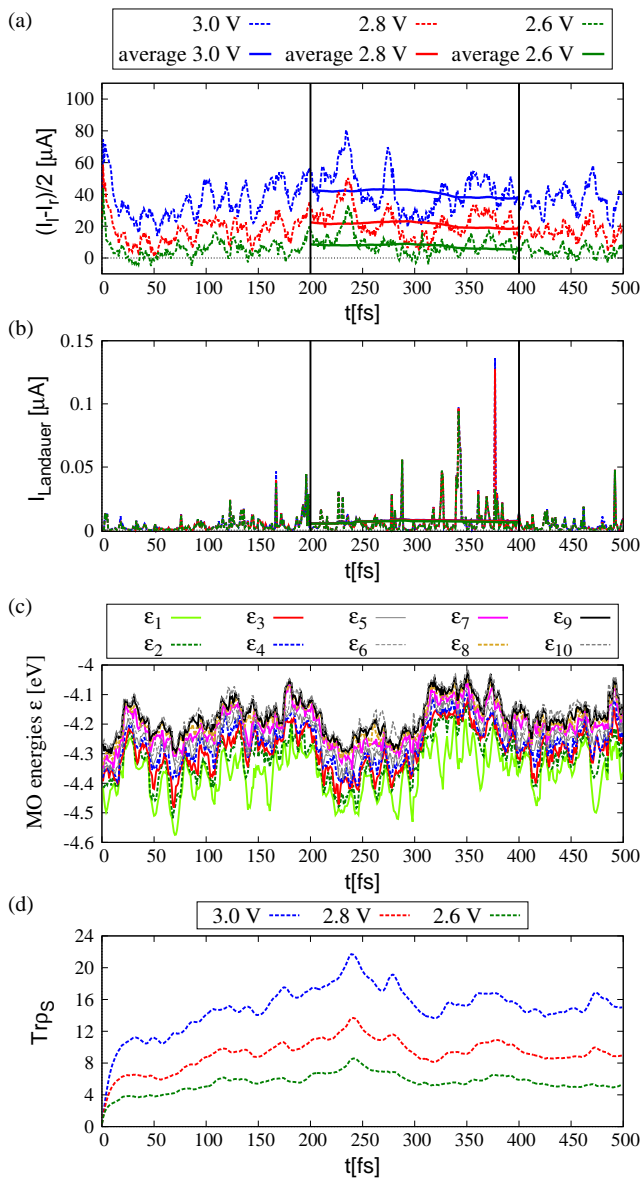}
  \caption{
  Current obtained with the TD-NEGF method (a) and the Landauer approach (b) for bias voltage 2.6, 2.8 and 3.0 V. 
  The dashed lines are the absolute currents, while the bold lines refer to the current averaged over a symmetric time span of 200 fs.
  (c) Time-resolved MO energies $\epsilon_1$-$\epsilon_{10}$, where $\epsilon_1$ corresponds to the LUMO of the SAM.
  (d) $\overline N=\text{Tr}\rho_S$, representing the number of electrons in the system for 2.6, 2.8 and 3.0 V. 
 The temperature is $T=300$ K.}
     \label{fig:005a}
\end{figure}

Fig.\ \ref{fig:005a} shows the results of the transport calculations over a simulation time of 500 fs for selected bias voltages. In addition to the TD-NEGF results in panel (a),
also the current obtained using the Landauer transport approach calculated at each snapshot of the MD simulation is depicted (b) as well as 
the time-evolution of the LUMO energies $\epsilon_1-\epsilon_{10}$ (c) and the total number of electrons in the SAM (d), given by $\overline N=\text{Tr}\rho_S$.
During the first 100 fs, the current exhibits pronounced changes, which is due to the fact that the simulation starts with an electronically unoccupied system far from steady state. This can also be seen in the evolution of the number of electrons in the SAM in panel (d), which reveals a rapid growth within the first 100 fs until a quasi steady state is reached.
After this transient period, 
the current oscillates with a frequency similar to that of the energy levels (panel (c)) around an average value, which increases with bias voltage.
Occasionally, pronounced fluctuations occur, such as the peaks in the current and the populations at times $\sim 250$ fs. These peak structures can be traced back to the fact that 
in this time interval, the energy levels (cf.\ panel (c)) are lower and significantly closer to the Fermi level. As a consequence, more states are located in the transport window, yielding higher currents and populations.

The current obtained with the simpler Landauer approach calculated for each snapshot of the MD trajectory, depicted in  panel
Fig.\ \ref{fig:005a} (b), is about 2-3 orders of magnitude lower than the TD-NEGF current. 
Peaks of larger current in the results of the Landauer approach (e.g.\ at $t=377$ fs) are caused by contributions of more delocalized states, such as the state shown in Fig.\ \ref{fig:004} (d),
which facilitate coherent transport processes.  
However, these peak values are still significantly lower than the TD-NEGF current.
It should be emphasized that the TD-NEGF approach provides the numerically exact result for the model considered.
As has been shown previously,\cite{popescu} the pronounced deviations of the Landauer approach are typical for systems with rapidly fluctuating electronic parameters; 
in particular systems where the timescales of the structural fluctuations are comparable to those of the charge transport processes.
While in the Landauer approach the current is calculated for static conformations and only depends on the corresponding fixed energy landscape, the TD-NEGF approach also describes transport processes during which the energy levels may change.
These processes are neglected within the Landauer approach and the current is therefore considerably underestimated. 

\begin{figure}[t]
\centering
  \includegraphics[scale=0.72]{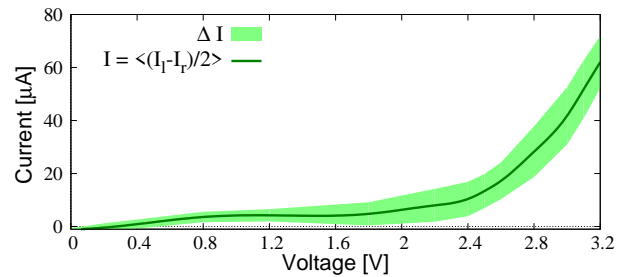}
       \caption{IV-characteristic obtained by averaging the TD-NEGF current over the time window 200-400 fs. The bold line depicts the time-averaged net current $I$, while the shaded area represents the fluctuations, given by the standard deviation $\Delta I$.}
          \label{fig:006}
\end{figure}

Averaging the TD-NEGF currents over a time range of 200 fs, an IV-characteristic is obtained as shown in Fig.\ \ref{fig:006}.  
The current increases first to a small plateau value for bias voltages in the range $0.8-2.2$ V and then to significantly larger values for higher voltages. This characteristics can be rationalized by the spectrum and character of the energy levels of the SAM. For bias voltages in the range $\approx 0.8-2.2$ V, only the lower unoccupied orbitals $\epsilon_1-\epsilon_{30}$ contribute to resonant transport. These orbitals are strongly localized (cf.\ Fig.\ \ref{fig:004}(b)), resulting in low currents. For larger voltages ($\ge$ 2.2 V), more delocalized MOs with stronger coupling to the electrodes (cf.\ Fig.\ \ref{fig:004}(c,d)) enter the transport window resulting in a pronounced increase of the current. 
At the onset of these two transport regimes the IV-characteristics exhibits pronounced broadening, which is caused by both thermal fluctuations and the coupling to the electrodes $\Gamma$.
%


\textit{Conclusions}. 
We have presented a multiscale method to study charge transport in organic semiconductors, which combines MD simulations, electronic structure calculation and TD-NEGF transport theory. 
The methodology is based on the approach developed by Popescu et al.\ \cite{popescu,Kubar16} for molecular junctions and extends it for applications to significantly larger systems.

As a representative example for organic semiconductors, we have applied the methodology to investigate charge transport  in C$_{60}$-based SAMs, which are used in SAMFET devices.
The results show that in these systems thermal fluctuations of the molecular structures induce pronounced rapid fluctuations of the electronic structure.
The influence of such rapid fluctuations on charge transport is correctly described within the TD-NEGF scheme employed, but is missing in simpler approaches that use Landauer theory for snapshots. As a result, Landauer theory predicts too low currents for the system investigated, in agreement with previous studies for model systems and charge transport in DNA.\cite{popescu,Kubar16}

In future work, the methodology presented here can be extended further by including the coupling to electrodes explicitly in the transport simulations\cite{Prucker2013} and electronic-vibrational coupling\cite{Wang2013} as well as electric field effects on the electronic structure and the back action of the electronic structure on the MD simulation.\cite{Kubar16} This may pave the way for a comprehensive treatment of charge transport in organic semiconductors without a priori assumptions about the dominant transport mechanism.

\textit{Acknowledgments}
We thank B.\ Popescu and U.\ {Kleinekath\"ofer} for helpful discussions.
This work has been supported by the
the Deutsche Forschungsgemeinschaft (DFG) through the Cluster
of Excellence 'Engineering of Advanced Materials' (EAM) and
SFB 953. Generous allocation of
computing time at the computing centers in Erlangen (RRZE),
Munich (LRZ), and J\"ulich (JSC) is gratefully acknowledged.

\end{document}